\documentclass[prl,reprint,twocolumn,showpacs,preprintnumbers,amsmath,amssymb,amsfonts,superscriptaddress]{revtex4}

\usepackage{graphicx}
\usepackage{dcolumn}
\usepackage{bm}
\usepackage{color}
\usepackage{amsmath,amsthm,amssymb,bm}
\usepackage{mathrsfs}
\usepackage{comment}

\newcommand{\Rb}{^{\text{87}}\text{Rb}}

\newcommand{\annih}[1]{\hat{\psi}_{#1}}
\newcommand{\creat}[1]{\hat{\psi}^\dagger_{#1}}

\newcommand{\R}{\mathbf{r}}

\newcommand{\p}{\mathbf{p}}

\newcommand{\eps}[1]{\epsilon^0_{#1}}

\newcommand{\integral}{\int\text{d}\mathbf{r}}

\graphicspath{%
    {converted_graphics/}
    {/}
    {Figures/}
}

\begin{document}

\preprint{APS/123-QED}


\title{Quantum Mass Acquisition in Spinor Bose-Einstein Condensates}
\author{Nguyen Thanh Phuc}
\affiliation{RIKEN Center for Emergent Matter Science (CEMS), Wako, Saitama 351-0198, Japan}
\author{Yuki Kawaguchi}
\affiliation{Department of Applied Physics, University of Tokyo, 7-3-1 Hongo, Bunkyo-ku, Tokyo 113-8656, Japan}
\author{Masahito Ueda}
\affiliation{Department of Physics, University of Tokyo, 7-3-1 Hongo, Bunkyo-ku, Tokyo 113-0033, Japan}
\affiliation{RIKEN Center for Emergent Matter Science (CEMS), Wako, Saitama 351-0198, Japan}
\date{\today}

\begin{abstract}
Quantum mass acquisition, in which a massless (quasi-)particle becomes massive due to quantum corrections, is predicted to occur in several subfields of physics. However, its experimental observation remains elusive since the emergent energy gap is too small. We show that a spinor Bose-Einstein condensate is an excellent candidate for the observation of such a peculiar phenomenon as the energy gap turns out to be two orders of magnitude larger than the zero-point energy. This extraordinarily large energy gap is a consequence of the dynamical  instability. The propagation velocity of the resultant massive excitation mode is found to be decreased as opposed to phonons.
\end{abstract}

\pacs{03.75.Kk,03.75.Mn,67.85.Jk}

\maketitle

\textit{Introduction}. Since the first realizations of the superfluid-Mott insulator phase transition in systems of bosons~\cite{Greiner02} and fermions~\cite{Jordens08}, systems of ultracold atoms have been regarded as an ideal quantum simulator of condensed-matter physics. Recently, there have been proposals to use ultracold atoms for the study of lattice gauge theories~\cite{Banerjee13, Zohar13}. Moreover, since ultracold atoms can be manipulated and measured with unprecedented precision, they are especially suited to investigate purely quantum-mechanical effects, which are very small and thus require high-resolution probes. Noticeable examples are quantum anomaly and vacuum alignment phenomena. The former is a weak violation of the original symmetry in the classical counterpart of the theory due to regularization, as proposed in two-dimensional Bose and Fermi gases~\cite{Olshanii10, Hofmann12}. The latter indicates the lift of the degeneracy in the ground-state manifold due to quantum fluctuations, as in the nematic phase of spin-2 BECs~\cite{Song07, Turner07}. 

In this Letter, we show that spinor Bose-Einstein condensates (BECs) offer an excellent playground for the study of quantum mass acquisition. It occurs as a purely quantum-mechanical effect in which a massless particle or quasiparticle becomes massive as a consequence of higher-order quantum corrections. This peculiar type of particles are called quasi-Nambu-Goldstone (qNG) bosons, which are gapless exciations that do not originate from spontaneous symmetry breaking. The qNG bosons have been a vital ingredient in high-energy physics~\cite{Weinberg-book, Kugo84, Bando88, Weinberg72, Georgi75}. They behave like Goldstone bosons at the zeroth order but acquire energy gaps due to higher-order corrections~\cite{Coleman73}. The qNG mode has also been predicted to appear in superfluid $^3$He~\cite{Volovik-book}, spin-1 color superconductors~\cite{Pang11}, spinor BECs~\cite{Uchino10}, and pyrochlore magnets~\cite{Ross14}. Despite its ubiquitous nature across different subfields of physics, the experimental observation of quantum mass acquisition has remained elusive because the emergent energy gap is too small to be distinguished from other secondary effects.

It is generally held that zero-point fluctuations should play a crucial role in determining the energy gap of the qNG mode~\cite{Uchino10}. However, it turns out that in contrast to other systems, in spinor BECs the zero-point energy does not set the energy scale of the gap of the qNG mode because the mean-field state, on which the zero-point energy is calculated, turns out to be dynamically unstable once quantum corrections beyond the Bogoliubov approximation are taken into account~\cite{Phuc13b}. In fact, by using the spinor Beliaev theory~\cite{Beliaev1, Beliaev2, Phuc13a, Phuc13b}, we analytically derive the emergent energy gap of the qNG mode in spin-2 BEC as a function of the fundamental interaction parameters and find that the energy gap is two orders of magnitude larger than the zero-point energy; therefore, the ground state is much more robust than previously imagined. This unexpectedly large energy gap gives us a great hope for the long-sought qNG mode to be observed experimentally. We also show that the propagation velocity of the qNG mode is decreased by fluctuations in particle-number density as opposed to phonons. 

\begin{figure}[tbp] 
  \centering
  \includegraphics[width=2.5in,keepaspectratio]{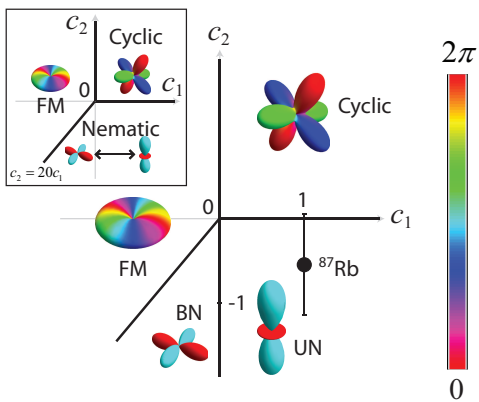}
  \caption{(Color online) Ground-state phase diagram of spin-2 BECs with quantum fluctuations. The interactions $c_1$ and $c_2$ are shown in units of $4\pi\hbar^2a_\mathrm{B}/M$, where $a_\mathrm{B}$ is the Bohr radius. The inset shows the corresponding mean-field phase diagram. The zero-point fluctuations lift the degeneracy in the nematic phase, selecting the uniaxial-nematic (UN) (biaxial-nematic (BN)) phase as the true ground state for $c_1>0$ ($c_1<0$). They also shift the ferromagnetic (FM)-BN and cyclic-UN phase boundaries~\cite{Phuc13b}, but the shifts are too small to be seen in the above scales of the axes. The values of $c_1$ and $c_2$ for $\Rb$ determined from the spin-exchange dynamics experiments~\cite{Widera06} and their error bars imply that the ground state is likely to be the UN phase. The spherical harmonic representation of the spinor order parameter is also displayed with the color showing the phase according to the color gauge on the right.}
  \label{fig: ground-state phase diagram}
\end{figure}

\textit{System}. The interaction energy of an ultracold dilute gas of spin-2 atoms is given in the second quantization by~\cite{Ciobanu00, Koashi00}
\begin{align}
\hat{V}=\frac{1}{2}\integral\Big[&c_0:\hat{n}^2:+c_1:\hat{\mathbf{F}}^2:+c_2:\hat{A}_{00}^\dagger\hat{A}_{00}:\Big],
\label{eq: Interaction Hamiltonian}
\end{align}
where $::$ denotes the normal ordering of operators. Here, $\hat{n}\equiv\sum_j\creat{j}(\R)\annih{j}(\R)$, $\hat{\mathbf{F}}\equiv\sum_{i,j}\creat{i}(\R)(\mathbf{f})_{ij}\annih{j}(\R)$, and $\hat{A}_{00}\equiv(1/\sqrt{5})\sum_j (-1)^{-j}\annih{j}(\R)\annih{-j}(\R)$ are the particle-number density, spin density, and spin-singlet pair amplitude operators, respectively, where $\annih{j}(\R)$ is the field operator of an atom in magnetic sublevel $j$ ($j=2,1,\cdots,-2$) at position $\R$, and $(\mathbf{f})_{ij}$ denotes the $ij$-component of the spin-2 matrix vector. The coefficients $c_0, c_1$, and $c_2$ are related to the $s$-wave scattering lengths $a_\mathcal{F}$ ($\mathcal{F}=0,2,4$) of the total spin-$\mathcal{F}$ channel by $c_0=4\pi\hbar^2(4a_2+3a_4)/(7M)$, $c_1=4\pi\hbar^2(a_4-a_2)/(7M)$, and $c_2=4\pi\hbar^2(7a_0-10a_2+3a_4)/(7M)$. For spin-2 $\Rb$ atoms, the magnitude of the spin-independent interaction is given by $c_0/(4\pi\hbar^2/M)\simeq a_{0, 2, 4}\simeq 100 a_\mathrm{B}$, where $a_\mathrm{B}$ is the Bohr radius. According to the spin-exchange dynamics measurement~\cite{Widera06}, the value of $c_1$ is accurately determined to be $c_1/(4\pi\hbar^2a_\mathrm{B}/M)=0.99\pm0.06$, whereas that of $c_2$ suffers a large error bar: $c_2/(4\pi\hbar^2a_\mathrm{B}/M)=-0.53\pm0.58$. As shown in Fig.~\ref{fig: ground-state phase diagram}, the negative median of $c_2$ implies that the ground state of $\Rb$ BEC is likely to be uniaxial-nematic (UN). A numerical calculation of $c_2$ for $\Rb$ also supports the high likelihood of the ground state being the UN phase~\cite{Klausen01}.

The three interactions in Eq.~\eqref{eq: Interaction Hamiltonian} with coupling constants $c_0$, $c_1$, and $c_2$ have SU(5), SO(3), and SO(5) symmetries, respectively~\cite{Uchino10b}; the symmetry of the total Hamiltonian is therefore SO(3). As a result, the maximum number of continuous symmetries that can be spontaneously broken is three, leading to three Nambu-Golstone excitations (one phonon and two magnon modes). The Bogoliubov spectrum of the nematic phase in spin-2 BECs shows a total of five gapless excitations~\cite{Ueda02}. The two extra gapless modes that do not stem from spontaneous symmetry breaking are the qNG modes. The appearance of the two extra gapless excitations can be understood by noting that the mean-field ground-state manifold of the nematic phase has an SO(5) symmetry which is larger than that of the Hamiltonian. In fact, the UN, biaxial-nematic (BN), and dihedral-2 phases belonging to this manifold can be transformed to each other by SO(5) rotations in spin space~\cite{Uchino10}. However, quantum fluctuations lift the degeneracy in the nematic phase and select the UN phase as the true ground state of $\Rb$~\cite{Song07, Turner07} (see Fig.~\ref{fig: ground-state phase diagram}). As a result, the qNG modes are expected to acquire a finite energy gap~\cite{Uchino10}, i.e., the phenomenon of quantum mass acquisition~\cite{Coleman73}. 

\textit{Emergent energy gap}. Since the emergent energy gap of the qNG modes results from the lift of the degeneracy caused by quantum fluctuations, it might be thought that the energy scale of the gap is determined by the zero-point energy per particle $E_\mathrm{zp}\equiv E^\mathrm{BN}-E^\mathrm{UN}\simeq 6.4c_1n(c_1/c_0)^{3/2}\sqrt{na^3}$, where $E^\mathrm{BN}$ and $E^\mathrm{UN}$ are the energies of the BN and UN phases with the Lee-Huang-Yang corrections~\cite{Uchino10b, Kawaguchi12}. This energy scale is too small in typical ultracold atomic experiments~\cite{Leanhardt03}, and therefore it seems impossible to probe the qNG modes. However, we find that the BN phase is dynamically unstable as its excitation energy involves a nonzero imaginary part due to beyond-Bogoliubov quantum corrections~\cite{Phuc13b}, and therefore the zero-point energy plays no role in the mass acquisition mechanism. In fact, the energy gap of the qNG modes will be shown to be much larger than previously imagined.  

To directly evaluate the magnitude of the emergent energy gap of the qNG modes, we use the spinor Beliaev theory~\cite{Beliaev1, Beliaev2, Phuc13a, Phuc13b}. We start from the Dyson equation~\cite{Supp, Fetter-Walecka}
\begin{align}
G^{\alpha\beta}_{jj'}(p)=(G^0)^{\alpha\beta}_{jj'}(p)+(G^0)^{\alpha\gamma}_{jm}\Sigma^{\gamma\delta}_{mm'}(p)G^{\delta\beta}_{m'j'}(p),
\label{eq: Dyson equation}
\end{align}
where $p\equiv(\omega,\p)$ denotes a frequency-wavenumber four-vector, and $G$, $G^0$, and $\Sigma$ represent the interacting Green's function, the noninteracting Green's function, and the self-energy, respectively; they are $10\times 10$ matrices with $j,j',m,m'=2,1,\cdots,-2$ labeling the magnetic sublevels and the values of $\alpha,\beta,\gamma,\delta$ indicating the normal (11,22) and anomalous (12,21) components. The noninteracting Green's function is given by $G^0_{jj}(p)=[\omega-(\eps{\p}-\mu)/\hbar+i\eta]^{-1}$, where $\eps{\p}\equiv \hbar^2\p^2/(2M)$ with $M$ being the atomic mass, $\mu$ is the chemical potential, and $\eta$ is an infinitesimal positive number. For the UN phase, the condensate atoms occupy only the $m_F=0$ state. The corresponding spinor order parameter is $\vec{\xi}=(0,0,1,0,0)^\mathrm{T}$. Since the UN phase possesses the time-reversal and space-inversion symmetries, there is a twofold degeneracy in the excitation spectrum $\omega_{j,\p}=\omega_{-j,\p}$.

For a weakly interacting dilute Bose gas, it is appropriate to make expansions of $\Sigma$ and $\mu$ with respect to the small dimensionless parameter $na^3\ll 1$, where $n$ is the total atomic number density and $a\equiv (4a_2+3a_4)/7$ is the average $s$-wave scattering length ($c_0=4\pi\hbar^2a/M$). These expansions are represented by the sums of Feynman diagrams as $\Sigma=\sum_{n=1}^\infty \Sigma^{(n)}$ and $\mu=\sum_{n=1}^\infty \mu^{(n)}$, where $\Sigma^{(n)}$ and $\mu^{(n)}$ are the contributions to the self-energy and the chemical potential from the $n$th-order Feynman diagrams~\cite{Phuc13a, Phuc13b}. The Bogoliubov and Beliaev theories include the Feynman diagrams up to the first and second orders, respectively. The second-order diagrams involve virtual excitations of the condensate, i.e., quantum fluctuations, which are absent in the first-order diagrams~\cite{Supp}. It is these quantum fluctuations that give rise to the energy gap of the qNG modes as shown below. 

In the first-order calculation, the poles of the Green's function reproduce the Bogoliubov spectra~\cite{Supp}: $\hbar\omega^{(1)}_{\pm2,\p}=\sqrt{\eps{\p}(\eps{\p}-2c_2n/5)}$, $\hbar\omega^{(1)}_{\pm1,\p}=\sqrt{\eps{\p}\left[\eps{\p}+2(3c_1-c_2/5)n\right]}$, and $\hbar\omega^{(1)}_{0,\p}=\sqrt{\eps{\p}\left[\eps{\p}+2(c_0+c_2/5)n\right]}$. All of the five elementary excitations are gapless. Specifically, $\omega^{(1)}_{0,\p}$ and $\omega^{(1)}_{\pm1,\p}$ are the spectra of one phonon and two magnons, respectively, which arise from spontaneous breaking of the gauge and spin-rotational symmetries, while the remaining two gapless quadrupolar (nematic) excitations $\omega^{(1)}_{\pm2,\p}$ are the qNG modes.

\begin{figure}[htbp] 
  \centering
  \includegraphics[width=2.5in,keepaspectratio]{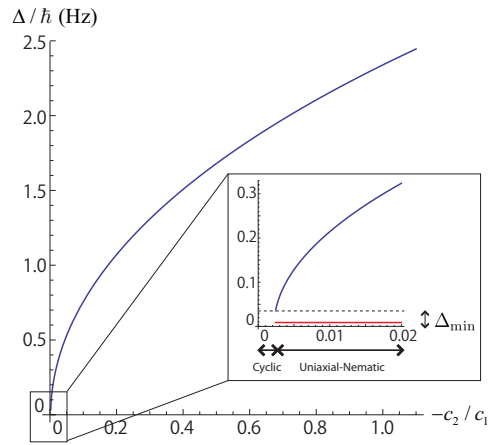}
  \caption{(Color online) Emergent energy gap of the quasi-Nambu-Goldstone modes in the uniaxial-nematic phase of the spin-2 $\Rb$ BEC (blue curve) as a function of the relative strength of the spin-dependent interactions. Here we use the same parameters of $\Rb$ as Fig.~\ref{fig: ground-state phase diagram} with the atomic number density $n=4\times10^{14}\,\text{cm}^{-3}$~\cite{Chang04, Schmaljohann04}. The inset magnifies the region of $-c_2\ll c_1$. There exists a positive lower bound $\Delta_\mathrm{min}$ for the energy gap of qNG modes. The zero-point energy per particle (red line) is shown for comparison. Note that the UN-cyclic phase boundary is shifted from its mean-field counterpart at $c_2=0$ due to the zero-point energy~\cite{Phuc13b}.}
  \label{fig: quasi-NG mode's energy gap}
\end{figure}

By adding the contributions to $\Sigma$ and $\mu$ from the second-order Feynman diagrams,  we obtain the emergent energy gap $\Delta\equiv\hbar\omega_{\pm2,\p=\boldsymbol{0}}$ of the qNG modes as~\cite{Supp}
\begin{align}
\left(\frac{\Delta}{c_1n}\right)^2=f(x)\left[x-\frac{64}{\sqrt{3\pi}}\left(\frac{c_1}{c_0}\right)^\frac{3}{2}\sqrt{na^3}\right]\left(\frac{c_1}{c_0}\right)^\frac{3}{2}\sqrt{na^3},
\label{eq: energy gap of quasi-NG modes}
\end{align}
where $x\equiv-c_2/(15c_1)$ and $f(x)\equiv 64\sqrt{108/\pi} \Big[(1+x)^\frac{5}{2}-$\\$4x(1+x)^\frac{3}{2}+2x^\frac{3}{2}(1+x)+3x^2(1+x)^\frac{1}{2}-2x^\frac{5}{2}\Big]$. Using the parameters of $\Rb$ with $n=4\times10^{14}\,\text{cm}^{-3}$ as in the experiments in Refs~\cite{Chang04, Schmaljohann04}, we plot the energy gap as a function of $-c_2/c_1$ over the uncertainty range of the interaction $c_2$~\cite{Widera06} in Fig.~\ref{fig: quasi-NG mode's energy gap}. Notice that $\Delta$ strongly depends on the relative strength of the spin-dependent interactions, and thus its measurement can provide us with useful information about the coupling strength $c_2$ that has yet to be precisely determined. In particular, the value of $\Delta$ at the UN-cyclic phase boundary $c_2^\mathrm{UN-CL}=-342c_1(c_1/c_0)^{3/2}\sqrt{na^3}$~\cite{Phuc13b} gives the lower bound for the energy gap: $\Delta_\mathrm{min}\simeq27.06\,c_1n\left(\frac{c_1}{c_0}\right)^\frac{3}{2}\sqrt{na^3}>0$. This result shows that the qNG modes acquire a finite energy gap; i.e., they become massive. At $c_2=-c_1$, the energy gap reduces to $\Delta\simeq 48/(5\sqrt[4]{5\pi})c_1n(c_1/c_0)^{3/4}\sqrt[4]{na^3}$, the magnitude of which is as large as $\Delta/\hbar\simeq 2.3$ Hz. We note that since $c_1/c_0<1$ and $na^3\ll 1$, the obtained energy gap is two orders of magnitude larger than the zero-point energy per particle $E_\mathrm{zp}\simeq6.4c_1n(c_1/c_0)^{3/2}\sqrt{na^3}\simeq \hbar\times$0.008 Hz (see Fig.~\ref{fig: quasi-NG mode's energy gap}). The obtained large energy gap implies that the ground state of the UN phase is much more robust than previously imagined. 

Since the qNG modes are gapless at the Bogoliubov level, the energy gap obtained by the Beliaev theory is the leading-order term in the asymptotic expansion of the excitation energy with respect to the dimensionless parameter $na^3\ll 1$, and higher-order terms should be negligibly small. In fact, although the emergent energy gap is two orders of magnitude larger than the zero-point energy, it is still small compared with the mean-field interaction energies, which ensures the validity of the perturbative expansion. For the parameters used in Fig.~\ref{fig: quasi-NG mode's energy gap}, the mean-field spin-dependent interaction energy is approximately $\hbar\times$190 Hz at $c_2=-c_1$, which is much larger than the obtained energy gap. Moreover, for a spinless (scalar) BEC, it has been justified by both experiments and quantum Monte-Carlo simulations that higher-order terms beyond the Beliaev theory are negligibly small for the parameters of typical ultracold atom experiments~\cite{Navon11}.

The energy gap of the qNG modes can be directly measured by the phase-contrast imaging. Indeed, the qNG mode with the lowest energy can be created in a finite system by uniformly and coherently transferring a fraction of atoms from the $|m_F=0\rangle$ to $|m_F=\pm2\rangle$ hyperfine states. The relative phase between those atoms then oscillates with a frequency given by the energy gap of the qNG modes. Since the phase-contrast imaging signal depends on the relative phase between $|m_F=0\rangle$ and $|m_F=\pm2\rangle$ atoms, which reflects the nematicity of the spinor condensate~\cite{Carusotto04, Higbie05}, the energy gap of the qNG modes can be directly obtained from the oscillating component of the probe signal. A similar method has been used to measure the dipolar energy gap of magnons in the spin-1 ferromagnetic phase~\cite{Marti14}. In our case, however, the effect of the magnetic dipole interaction on the energy gap of the qNG modes is strongly suppressed since the magnetization of the condensate remains zero as long as atoms are not excited to the $|m_F=\pm1\rangle$ states. It should also be noted that the finite size of a trapped BEC makes the fluctuation-induced energy gap of the qNG modes larger, rather than smears it out, due to the enhancement of quantum fluctuations in a confined system. Therefore, the right-hand side of Eq.~\eqref{eq: energy gap of quasi-NG modes} gives the lower bound for the energy gap in a trapped condensate.

Once the qNG modes acquire a mass, the $\mathbb{Z}_2$ vortex, a topological structure, is stabilized and might be observed in experiments. Around this vortex, the symmetry axis of the UN phase rotates by $\pi$ with the order parameter far from the vortex core given by $\vec{\psi}_{\mathbb{Z}_2}=\sqrt{n_0}(\sqrt{6}e^{i\phi}/4,0,-1/2,0,\sqrt{6}e^{-i\phi}/4)^\mathrm{T}$, where $\phi$ denotes the azimuthal angle in real space~\cite{Kawaguchi12}. The $w=\pm1$ phase winding numbers for the $m_F=\pm2$ spin components can be created by transferring the angular momentum between a Laguerre-Gaussian beam and atoms~\cite{Leslie09}. Inside the vortex core, the order parameter deviates from that of the UN phase but lies within the manifold of nematic phases. Hence, the core size is determined by the balance between the kinetic energy and the energy difference between the two phases, which is given by the energy gap. The estimated energy gap $\Delta/\hbar\simeq 2.3$ Hz corresponds to the vortex core of the order of tens of microns, which is slightly larger than the typical Thomas-Fermi length of a BEC. The vortex core can, however, be made smaller in, for example, a highly oblate condensate since fluctuations and consequently the energy gap would be enhanced in a lower dimension. Notice that even when the energy gap of the qNG modes is much smaller than the temperature of the condensate, the vortex still remains stable as a consequence of Bose enhancement. A similar coherence effect has been observed in the superfluid $^3$He~\cite{Blaauwgeers00}.

\textit{Propagation of quasiparticles}. In addition to the emergent energy gap of the qNG modes, quantum fluctuations also give a correction to the propagation velocity of these quasiparticles. In the low-momentum regime, the Bogoliubov dispersion relation of the qNG modes is linear: $\omega^{(1)}_{\pm2,\p}=v^{(1)}_\mathrm{qNG}|\p|$ with the first-order propagation velocity given by $v^{(1)}_\mathrm{qNG}=\sqrt{|c_2|n/(5M)}$. Since $c_0\gg c_1, |c_2|$ for the spin-2 $\Rb$ BEC, concerning the modification of the propagation velocity, we can concentrate on the effect of fluctuations in particle-number density and ignore that of spin-density fluctuations. In the momentum regime satisfying $\Delta\ll \eps{\p}\ll |c_2|n$ where the modified dispersion relation is linear, the spectrum of the qNG modes can be expressed as $\omega^{(2)}_{\pm2,\p}\simeq\,\omega^{(1)}_{\pm2,\p}+\left[\Sigma^{11(2)}_{22}(p)-\Sigma^{11(2)}_{22}(-p)\right]/2$. Here we ignore terms involving small factors of $\Delta/\eps{\p}$ and $\eps{\p}/(|c_2|n)$. By summing all the contributions to $\Sigma^{11}_{22}$ from the second-order Feynman diagrams, we obtain the second-order dispersion relation of the qNG modes $\omega^{(2)}_{\pm2,\p}=v^{(2)}_\mathrm{qNG}|\p|$ with the modified propagation velocity $v^{(2)}_\mathrm{qNG}=\left(1-4\sqrt{na^3/\pi}\right)\sqrt{|c_2|n/(5M)}$~\cite{Supp}. From the expressions of $v^{(1)}_\mathrm{qNG}$ and $v^{(2)}_\mathrm{qNG}$, we find that the propagation of the qNG modes is hindered by their interactions with noncondensed atoms. A similar phenomenon occurs with magnons in spin-1 BECs which has been predicted~\cite{Phuc13a} and observed~\cite{Marti14}. This can be understood by noting that the qNG modes in spin-2 BECs represent the spatially periodic modulations of spin nematicity which are uncorrelated with fluctuations in particle-number-density, thus leading to the resistance. In contrast, the propagation velocity of phonons is enhanced by quantum fluctuations~\cite{Beliaev1}. The restoring force resulting from the inhomogeneity brought about by the density correlation at zero temperature makes the system more rigid with a smaller compressibility compared with a homogeneous state, which results in a larger sound velocity. 

\textit{Conclusion}. We have found that spinor Bose-Einstein condensates offer an excellent test bed for the quest of the quasi-Nambu-Goldstone mode and for the study of the phenomenon of quantum mass acquisition. The emergent energy gap of the qNG mode turns out to be two orders of magnitude larger than the zero-point energy, indicating much greater robustness of the ground state of the condensate than previously imagined. This extraordinarily large energy gap is a consequence of the dynamical instability in spinor condensates. The propagation velocity of the qNG quasiparticle is found to be decreased by fluctuations in particle-number density, which can be verified experimentally.   

\begin{acknowledgments}
We gratefully acknowledge G. Volovik for helpful discussions. N. T. P. would like to thank S. Uchino for useful correspondence. This work was supported by KAKENHI Grant No. 26287088 from the Japan Society for the Promotion of Science, and a Grant-in-Aid for Scientific Research on Innovation Areas \textquotedblleft Topological Quantum Phenomena" (KAKENHI Grant No. 22103005), and the Photon Frontier Network Program from MEXT of Japan. Y.K. acknowledges the financial support from KAKENHI Grant No.22740265 and Inoue Foundation. 
\end{acknowledgments}


\end{document}


\preprint{APS/123-QED}


\title{Supplemental Material for \textquotedblleft Quantum Mass Acquisition in Spinor Bose-Einstein Condensates"}
\author{Nguyen Thanh Phuc}
\affiliation{RIKEN Center for Emergent Matter Science (CEMS), Wako, Saitama 351-0198, Japan}
\author{Yuki Kawaguchi}
\affiliation{Department of Applied Physics, University of Tokyo, 7-3-1 Hongo, Bunkyo-ku, Tokyo 113-8656, Japan}
\author{Masahito Ueda}
\affiliation{Department of Physics, University of Tokyo, 7-3-1 Hongo, Bunkyo-ku, Tokyo 113-0033, Japan}
\affiliation{RIKEN Center for Emergent Matter Science (CEMS), Wako, Saitama 351-0198, Japan}
\date{\today}

\begin{abstract}
A detailed derivation of the emergent energy gap of the quasi-Nambu-Goldstone (qNG) modes in the spin-2 Bose-Einstein condensate is given. The formalism of the spinor Beliaev theory is introduced, followed by the evaluation of the second-order self-energies, from which the energy gap is calculated. Finally, the quantum correction to the propagation velocity of the qNG modes is evaluated.
\end{abstract}

\pacs{03.75.Kk,03.75.Mn,67.85.Jk}

\maketitle

\section{Spin-2 spinor Beliaev theory}
\label{sec: Spinor Beliaev theory: formalism}
\begin{figure}[tbp] 
  \centering
  \includegraphics[width=3in, keepaspectratio]{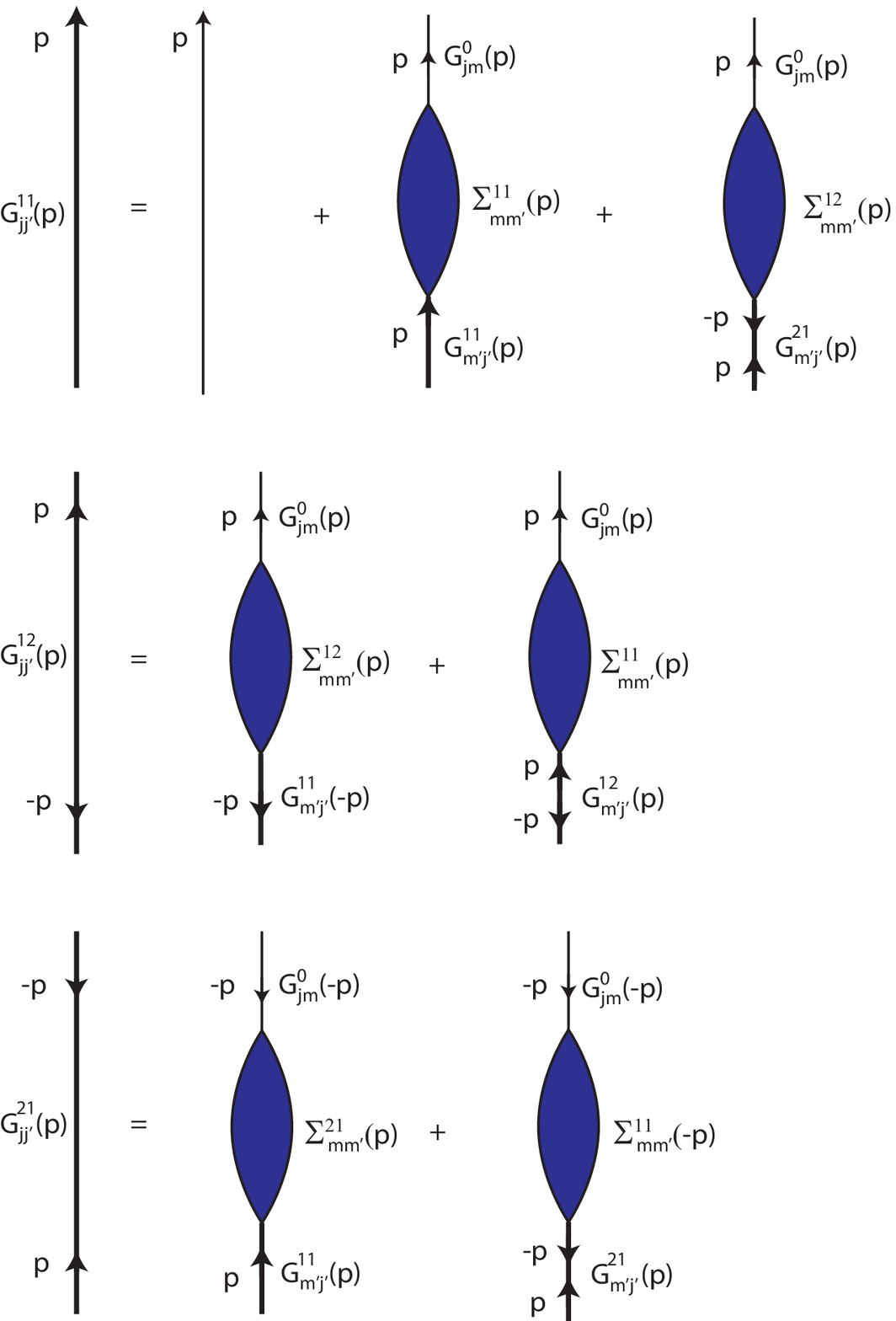}
  \caption{(Color online) Dyson's equations for the normal and anomalous components of the Green's function. The thick and thin lines, and the oval represent the interacting and noninteracting Green's functions, and the self-energy, respectively~\cite{Phuc13a}.}
  \label{fig: Dyson equation}
\end{figure}

We first develop the Beliaev theory for a spin-2 Bose-Einstein condensate (BEC), which is a Green's function approach that involves Feynman diagrams up to the second order~\cite{Beliaev1, Beliaev2}. It gives the excitation spectrum of the condensate at the next order beyond the Bogoliubov theory. The Dyson equation for the Green's function is given by Eq.~(2) in the main text and illustrated by the diagrams in Fig.~\ref{fig: Dyson equation}~\cite{Fetter-Walecka, Phuc13a, Phuc13b}. The normal and anomalous components of the Green's function as well as the self-energy describe the propagation of a single noncondensed particle and the creation or annihilation of a pair of noncondensed particles, respectively.

As discussed in the main text, the ground-state phase of the spin-2 $\Rb$ BEC is likely to be uniaxial-nematic (UN), where the representative order parameter is given by $\vec{\psi}=\sqrt{n_0}(0,0,1,0,0)^\mathrm{T}$ with $n_0$ being the number density of the condensate atoms. This means that the condensate atoms occupy only the magnetic sublevel $m_F=0$ of the $F=2$ hyperfine-spin manifold. The matrix $\Sigma^{\alpha\beta}_{jj'}$ then become block diagonal as
\begin{widetext}
\begin{align}
\Sigma^{\alpha\beta}_{jj'}=
\begin{bmatrix}
\Sigma^{11}_{2,2}&0&0&0&0&0&0&0&0&\Sigma^{12}_{2,-2}\\
0&\Sigma^{11}_{1,1}&0&0&0&0&0&0&\Sigma^{12}_{1,-1}&0\\
0&0&\Sigma^{11}_{0,0}&0&0&0&0&\Sigma^{12}_{0,0}&0&0\\
0&0&0&\Sigma^{11}_{-1,-1}&0&0&\Sigma^{12}_{-1,1}&0&0&0\\
0&0&0&0&\Sigma^{11}_{-2,-2}&\Sigma^{12}_{-2,2}&0&0&0&0\\
0&0&0&0&\Sigma^{21}_{2,-2}&\Sigma^{22}_{2,2}&0&0&0&0\\
0&0&0&\Sigma^{21}_{1,-1}&0&0&\Sigma^{22}_{1,1}&0&0&0\\
0&0&\Sigma^{21}_{0,0}&0&0&0&0&\Sigma^{22}_{0,0}&0&0\\
0&\Sigma^{21}_{-1,1}&0&0&0&0&0&0&\Sigma^{22}_{-1,-1}&0\\
\Sigma^{21}_{-2,2}&0&0&0&0&0&0&0&0&\Sigma^{22}_{-2,-2}\\
\end{bmatrix}.
\label{eq: UN, Sigma matrix}
\end{align}
\end{widetext}
Here $\Sigma^{22}_{jj'}(p)\equiv\Sigma^{11}_{jj'}(-p)$ and $\Sigma^{12}_{jj'}=\Sigma^{21}_{jj'}$ because the corresponding diagrams are the same. This block-diagonal pattern of the self-energy can be understood as a consequence of the spin conservation in which the scattering of two condensate atoms with $m_F=0$ can only create a pair of noncondensed atoms with  $m_F=\pm j$. Substituting these self-energies into the Dyson equation, the Green's function turns out to be block diagonal as does the self-energy, and its nonzero components are given by
\begin{align}
G^{11}_{j,j}(p)=&\frac{-[G^0_{j,j}(-p)]^{-1}+\Sigma^{11}_{j,j}(-p)}{D_{j}},
\label{eq: Green function G11,jj for UN phase}
\end{align}
where
\begin{align}
D_{j}=&-[G^0_{j,j}(p)]^{-1}[G^0_{-j,-j}(-p)]^{-1}+\Sigma^{11}_{j,j}(p)[G^0_{-j,-j}(-p)]^{-1}\nonumber\\
&+\Sigma^{22}_{-j,-j}(p)[G^0_{j,j}(p)]^{-1}-\Sigma^{11}_{j,j}(p)\Sigma^{22}_{-j,-j}(p)\nonumber\\
&+\Sigma^{21}_{-j,j}(p)\Sigma^{12}_{j,-j}(p). 
\label{eq: denominator of the Green function G11,22 for UN phase}
\end{align}
The non-interacting Green's function is diagonal and its element is given by $G^0_{j,j}(p)=\left[\omega-(\eps{\p}-\mu)/\hbar+i\eta\right]^{-1}$, which is independent of the spin state $m_F=j$ in the absence of an external magnetic field. Here, $\eps{\p}\equiv \hbar^2\p^2/(2M)$ with $M$ being the atomic mass, $\mu$ is the chemical potential, and $\eta$ is an infinitesimal positive number. According to the Lehmann representation, the zeros of $D_j$ give the excitation energy spectrum, which is calculated for $\p=\boldsymbol{0}$ to be
\begin{align}
\omega_{j,\p=\boldsymbol{0}}=&\frac{\Sigma^{11}_{j,j}-\Sigma^{22}_{-j,-j}}{2}\pm\Bigg\{-\Sigma^{12}_{j,-j}\Sigma^{21}_{-j,j}\nonumber\\
&+\Bigg[-\frac{\mu}{\hbar}+\frac{\Sigma^{11}_{j,j}+\Sigma^{22}_{-j,-j}}{2}\Bigg]^2\Bigg\}^{1/2}.
\label{eq: solution to the Dyson equation: omega(j,p=0) for UN phase}
\end{align}
It should be noted that the self-energies on the right-hand side of Eq.~\eqref{eq: solution to the Dyson equation: omega(j,p=0) for UN phase} are functions of $\omega_{j,\p=\boldsymbol{0}}$, and the plus and minus signs in front of the square root result in a pair of poles of the Greens function. However, only the solution with the plus sign satisfies the canonical commutation relation for quasiparticles' operators. For the UN phase with a symmetric order parameter $\vec{\phi}=\sqrt{n_0}(0,0,1,0,0)^\mathrm{T}$, there is a symmetry between the $m_F=\pm j$ magnetic sublevels, leading to
\begin{align}
\Sigma^{11}_{j,j}=\Sigma^{11}_{-j,-j},\,\,&\Sigma^{22}_{j,j}=\Sigma^{22}_{-j,-j}, \label{eq: Sigma(jj)=Sigma(-j,-j)}\\
\Sigma^{12}_{j,-j}=\Sigma^{12}_{-j,j}=&\Sigma^{21}_{j,-j}=\Sigma^{21}_{-j,j},\label{eq: Sigma 12=Sigma21}\\
D_j=&D_{-j}.\label{eq: Dj=D-j}
\end{align}
Equation~\eqref{eq: Dj=D-j} reflects the twofold degeneracy in the excitation energies given by Eq.~\eqref{eq: solution to the Dyson equation: omega(j,p=0) for UN phase}.

\begin{figure}[htbp] 
  \centering
  \includegraphics[width=3in,keepaspectratio]{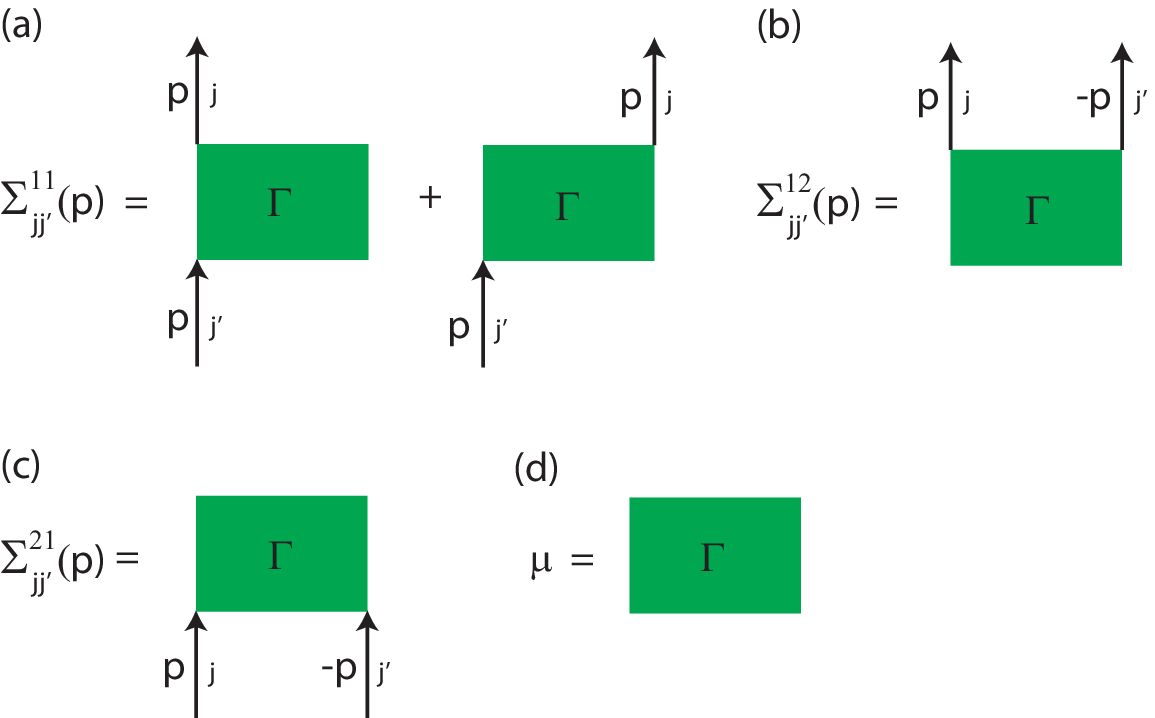}
  \caption{(Color online) First-order Feynman diagrams for the self-energies (a) $\Sigma^{11}_{jj'}(p)$, (b) $\Sigma^{12}_{jj'}(p)$, (c) $\Sigma^{21}_{jj'}(p)$, and (d) the chemical potential $\mu$. The two diagrams in (a) represent the Hartree (left) and Fock (right) interactions. Here $p\equiv(\p,\omega)$ and $j$ denote the frequency-momentum four-vector and the magnetic sublevel, respectively. The rectangles represent the $T$-matrices, where condensate particles are not explicitly shown. In (a), there are one condensate particle moving in and another moving out; in (b) and (c), there are two condensate particles moving in and two moving out, respectively; in (d), all four particles belong to the condensate~\cite{Phuc13a, Phuc13b}.}
  \label{fig: first-order Feynman diagrams}
\end{figure}

\begin{figure}[htbp] 
  \centering
  \includegraphics[width=3in,keepaspectratio]{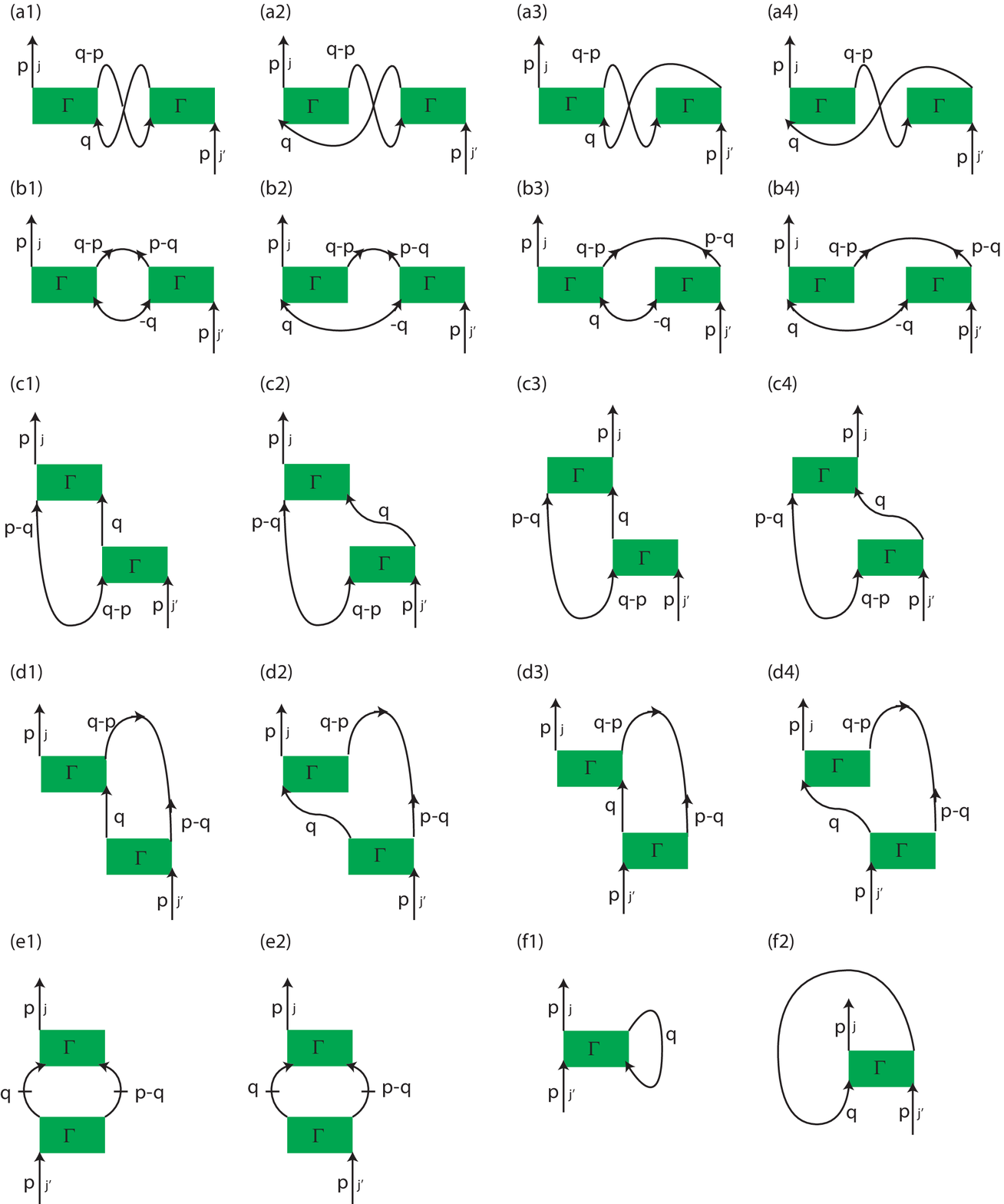}
  \caption{(Color online) Second-order Feynman diagrams for $\Sigma^{11}_{jj'}(p)$. The intermediate propagators are classified into three different categories, depending on the number of noncondensed atoms. They are represented by curves with one arrow ($\longrightarrow$), two out-pointing arrows ($\leftarrow$$\rightarrow$), and two in-pointing arrows ($\rightarrow$$\leftarrow$), which describe the first-order normal Green's function $G_{jj'}^{11}(p)$ and two anomalous Green's functions $G^{12}_{jj'}(p)$ and $G^{21}_{jj'}(p)$, respectively. Here, the two horizontal dashes in (e1) and (e2) indicate that the terms of noninteracting Green's functions are to be subtracted to avoid double counting of the contributions that have already been taken into account in the $T$-matrix and the first-order diagrams. As in Fig.~\ref{fig: first-order Feynman diagrams}, the condensate particles in (a1)--(e2) are not shown~\cite{Phuc13a, Phuc13b}.}
  \label{fig: second-order Feynman diagrams for Sigma11}
\end{figure}

\begin{figure}[htbp] 
  \centering
  \includegraphics[width=3in,keepaspectratio]{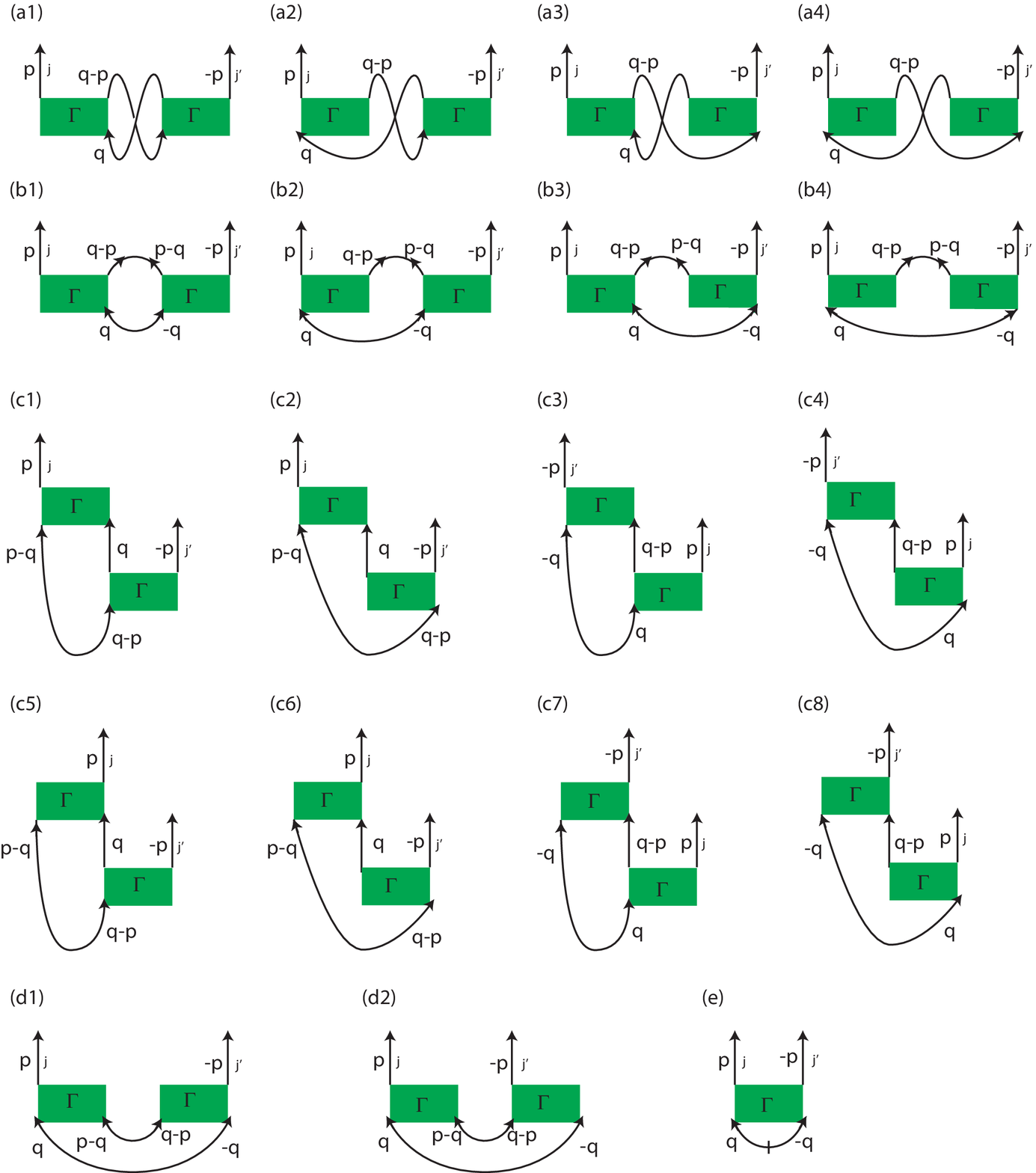}
  \caption{(Color online) Second-order Feynman diagrams for $\Sigma^{12}_{jj'}(p)$~\cite{Phuc13a, Phuc13b}.}
  \label{fig: second-order Feynman diagrams for Sigma12}
\end{figure}

\begin{figure}[htbp] 
  \centering
  \includegraphics[width=3in,keepaspectratio]{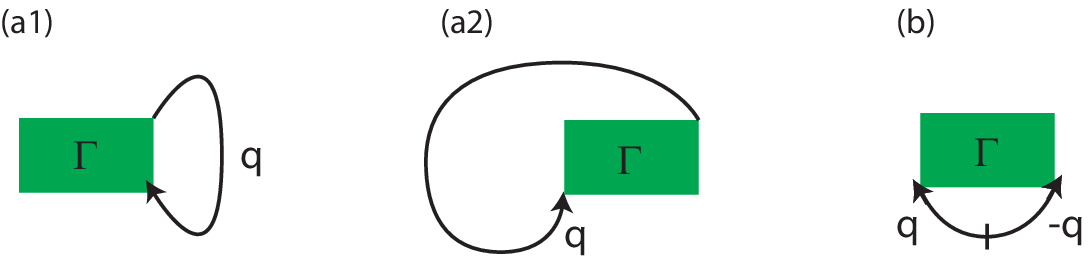}
  \caption{(Color online) Second-order Feynman diagrams for the chemical potential $\mu$~\cite{Phuc13a, Phuc13b}.}
  \label{fig: Second-order Feynman diagrams for the chemical potential mu}
\end{figure}

For a weakly interacting dilute Bose gase under consideration, it is appropriate to expand $\Sigma$ and $\mu$ with respect to the small dimensionless parameter $na^3\ll 1$, where $n$ is the total atom number density and $a\equiv (4a_2+3a_4)/7$ is the average s-wave scattering length ($c_0= 4\pi\hbar^2a/M$). These expansions are represented by the sums of Feynman diagrams,
\begin{subequations}
\label{eq: expansions of sigma and mu}
\begin{align}
\Sigma^{\alpha\beta}_{jj'}=&\sum_{n=1}^\infty \Sigma^{\alpha\beta(n)}_{jj'},\\
\mu=&\sum_{n=1}^\infty \mu^{(n)},
\end{align}
\end{subequations}
where $\Sigma^{\alpha\beta(n)}_{jj'}$ and $\mu^{(n)}$ are the contributions to the self-energy and the chemical potential from the $n$th-order Feynman diagrams. The Bogoliubov and Beliaev theories involve the Feynman diagrams up to the first order (Fig.~\ref{fig: first-order Feynman diagrams}) and the second order (Figs.~\ref{fig: second-order Feynman diagrams for Sigma11}--\ref{fig: Second-order Feynman diagrams for the chemical potential mu}), respectively. There appear virtual excitations, i.e., quantum fluctuations, of the condensate with momenta $q$ and $q-p$ in the second-order diagrams, which are absent in the first-order ones. It is these quantum fluctuations that give rise to the energy gap of the qNG modes as shown in the next section.

\section{Second-order self-energies}
\label{sec: Second-order self-energies}
In the absence of quantum fluctuations, the first-order self-energy and chemical potential are linear functions of the two-body interaction strengths: 
\begin{align}
\Sigma^{11(1)}_{\pm2,\pm2}=&c_0n_0, \label{eq: first-order Sigma22}\\
\Sigma^{11(1)}_{\pm1,\pm1}=&(c_0+3c_1)n_0,\\
\Sigma^{11(1)}_{0,0}=&2(c_0+c_2/5)n_0,\\
\Sigma^{12(1)}_{\pm2,\mp2}=&c_2n_0/5, \\
\Sigma^{12(1)}_{\pm1,\mp1}=&(3c_1-c_2/5)n_0,\\
\Sigma^{12(1)}_{0,0}=&(c_0+c_2/5)n_0,\\
\mu^{(1)}=&(c_0+c_2/5)n_0. \label{eq: first-order mu}
\end{align}
Substituting these into Eq.~\eqref{eq: solution to the Dyson equation: omega(j,p=0) for UN phase}, we obtain the following Bogoliubov spectra:
\begin{align}
\hbar\omega^{(1)}_{\pm2,\p}=&\sqrt{\eps{\p}(\eps{\p}-2c_2n/5)},\label{eq: first-order omega +-2}\\
\hbar\omega^{(1)}_{\pm1,\p}=&\sqrt{\eps{\p}\left[\eps{\p}+2(3c_1-c_2/5)n\right]},\\
\hbar\omega^{(1)}_{0,\p}=&\sqrt{\eps{\p}\left[\eps{\p}+2(c_0+c_2/5)n\right]}. \label{eq: first-order omega 0}
\end{align}
It is clear that all of the five elementary excitations are gapless. Specifically, $\omega^{(1)}_{0,\p}$ and $\omega^{(1)}_{\pm1,\p}$ are the spectra of one phonon and two magnons, respectively, which arise from spontaneous breaking of the gauge and spin-rotational symmetries, respectively, while the remaining two gapless quadrupolar (nematic) excitations $\omega^{(1)}_{\pm2,\p}$ are the qNG modes. In the following, we will show that by adding the contributions to the self-energy and the chemical potential from the second-order Feynman diagrams, a nonzero energy gap $\Delta\equiv \hbar\omega_{\pm2,\p=\boldsymbol{0}}$ emerges. 

By separating the contributions to $\Sigma$ and $\mu$ in Eq.~\eqref{eq: solution to the Dyson equation: omega(j,p=0) for UN phase} from the first- and second-order Feynman diagrams, we obtain the energy gap of the qNG modes up to the second order as
\begin{align}
\omega_{\pm2,\p=\boldsymbol{0}}=&\frac{\Sigma^{11(2)}_{22}-\Sigma^{22(2)}_{22}}{2}+\Bigg\{-\left[\frac{c_2n_0}{5\hbar}+\Sigma^{12(2)}_{2,-2}\right]^2\nonumber\\
&+\Bigg[-\frac{c_2n_0}{5\hbar}-\frac{\mu^{(2)}}{\hbar}+\frac{\Sigma^{11(2)}_{22}+\Sigma^{22(2)}_{22}}{2}\Bigg]^2\Bigg\}^{1/2},
\label{eq: solution to the Dyson equation: omega(2,p=0) for UN phase}
\end{align}
where the first-order self-energy and chemical potential given by Eqs.~\eqref{eq: first-order Sigma22}-\eqref{eq: first-order mu}, together with Eqs.~\eqref{eq: Sigma(jj)=Sigma(-j,-j)} and \eqref{eq: Sigma 12=Sigma21} were used.

By summing the contributions to $\Sigma^{11}_{22}$ from the second-order Feynman diagrams in Fig.~\ref{fig: second-order Feynman diagrams for Sigma11}, we obtain
\begin{widetext}
\begin{align}
\hbar\Sigma^{11(2)}_{22}(\omega,\p)=&n_0c_0^2\integralQ \Bigg[(A_{0,\K}+B_{0,\K}-2C_{0,\K})\left(\frac{A_{2,\Q}}{\hbar\left(\omega-\omega^{(1)}_{2,\Q}-\omega^{(1)}_{0,\K}\right)+i\eta}-\frac{B_{2,\Q}}{\hbar\left(\omega+\omega^{(1)}_{2,\Q}+\omega^{(1)}_{0,\K}\right)-i\eta}\right)\nonumber\\
&-\mathcal{P}\frac{1}{\eps{\p}-\eps{\Q}-\eps{\K}}\Bigg]+6n_0c_1^2\integralQ \Bigg[\frac{A_{1,\Q}(2A_{1,\K}+B_{1,\K}-4C_{1,\K})+C_{1,\Q}C_{1,\K}}{\hbar\left(\omega-\omega^{(1)}_{1,\Q}-\omega^{(1)}_{1,\K}\right)+i\eta}\nonumber\\
&-\frac{B_{1,\Q}(2B_{1,\K}+A_{1,\K}-4C_{1,\K})+C_{1,\Q}C_{1,\K}}{\hbar\left(\omega+\omega^{(1)}_{1,\Q}+\omega^{(1)}_{1,\K}\right)-i\eta}-2\mathcal{P}\frac{1}{\eps{\p}-\eps{\Q}-\eps{\K}}\Bigg]+\frac{4n_0c_0c_2}{5}\integralQ \nonumber\\
&\times\left[\frac{(C_{0,\Q}-A_{0,\Q})C_{2,\K}}{\hbar\left(\omega-\omega^{(1)}_{0,\Q}-\omega^{(1)}_{2,\K}\right)+i\eta}-\frac{(C_{0,\Q}-B_{0,\Q})C_{2,\K}}{\hbar\left(\omega+\omega^{(1)}_{0,\Q}+\omega^{(1)}_{2,\K}\right)-i\eta}\right]+\frac{4n_0c_2^2}{25}\integralQ\nonumber\\
&\times \left[\frac{A_{0,\Q}B_{2,\K}}{\hbar\left(\omega-\omega^{(1)}_{0,\Q}-\omega^{(1)}_{2,\K}\right)+i\eta}-\frac{B_{0,\Q}A_{2,\K}}{\hbar\left(\omega+\omega^{(1)}_{0,\Q}+\omega^{(1)}_{2,\K}\right)-i\eta}\right]+c_0\integralQ (3B_{2,\Q}+2B_{1,\Q}+B_{0,\Q})\nonumber\\
&+c_1\integralQ (2B_{1,\Q}+4B_{2,\Q})+\frac{2c_2}{5}\integralQ B_{2,\Q},
\label{eq: summing Feynman diagrams for Sigma 11(2), 22 for UN phase}
\end{align}
where $\K\equiv\Q-\p$ and $\mathcal{P}$ denotes the principle value of the integral. Here, the first-order, i.e., the Bogoliubov, excitation spectra of the UN phase are given by Eqs.~\eqref{eq: first-order omega +-2}--\eqref{eq: first-order omega 0}
and 
\begin{align} 
A_{2,\p}\equiv&\frac{\hbar\omega^{(1)}_{2,\p}+\eps{\p}-c_2n_0/5}{2\hbar\omega^{(1)}_{2,\p}},\\
B_{2,\p}\equiv&\frac{-\hbar\omega^{(1)}_{2,\p}+\eps{\p}-c_2n_0/5}{2\hbar\omega^{(1)}_{2,\p}},\\
C_{2,\p}\equiv&\frac{c_2n_0/5}{2\hbar\omega^{(1)}_{2,\p}},\\
A_{1,\p}\equiv&\frac{\hbar\omega^{(1)}_{1,\p}+\eps{\p}+(3c_1-c_2/5)n_0}{2\hbar\omega^{(1)}_{1,\p}},\\
B_{1,\p}\equiv&\frac{-\hbar\omega^{(1)}_{1,\p}+\eps{\p}+(3c_1-c_2/5)n_0}{2\hbar\omega^{(1)}_{1,\p}},\\
C_{1,\p}\equiv&\frac{(3c_1-c_2/5)n_0}{2\hbar\omega^{(1)}_{1,\K}},\\
A_{0,\p}\equiv&\frac{\hbar\omega^{(1)}_{0,\p}+\eps{\p}+(c_0+c_2/5)n_0}{2\hbar\omega^{(1)}_{0,\p}},\\
B_{0,\p}\equiv&\frac{-\hbar\omega^{(1)}_{0,\p}+\eps{\p}+(c_0+c_2/5)n_0}{2\hbar\omega^{(1)}_{0,\p}},\\
C_{0,\p}\equiv&\frac{(c_0+c_2/5)n_0}{2\hbar\omega^{(1)}_{0,\p}}.
\end{align}

The self-energy $\Sigma^{22(2)}_{22}$ is immediately obtained via $\Sigma^{22(2)}_{22}(\omega,\p)=\Sigma^{11(2)}_{22}(-\omega,-\p)$. Similarly, we obtain $\Sigma^{12(2)}_{2,-2}$ and $\mu^{(2)}$ by summing the contributions from the respective second-order Feynman diagrams in Figs.~\ref{fig: second-order Feynman diagrams for Sigma12} and \ref{fig: Second-order Feynman diagrams for the chemical potential mu}:
\begin{align}
\hbar\Sigma^{12(2)}_{2,-2}(\omega,\p)=&n_0c_0^2\integralQ C_{2,\Q}(2C_{0,\K}-A_{0,\K}-B_{0,\K})\left(\frac{1}{\hbar\left(\omega-\omega^{(1)}_{2,\Q}-\omega^{(1)}_{0,\K}\right)+i\eta}-\frac{1}{\hbar\left(\omega+\omega^{(1)}_{2,\Q}+\omega^{(1)}_{0,\K}\right)-i\eta}\right)\nonumber\\
&+6n_0c_1^2\integralQ \left[-C_{1,\K}(2A_{1,\Q}+2B_{1,\Q}-3C_{1,\Q})+A_{1,\Q}B_{1,\K}\right]\Bigg(\frac{1}{\hbar\left(\omega-\omega^{(1)}_{1,\Q}-\omega^{(1)}_{1,\K}\right)+i\eta}\nonumber\\
&-\frac{1}{\hbar\left(\omega+\omega^{(1)}_{1,\Q}+\omega^{(1)}_{1,\K}\right)-i\eta}\Bigg)+\frac{2n_0c_0c_2}{5}\integralQ \left[A_{2,\Q}B_{0,\K}+A_{0,\K}B_{2,\Q}-(A_{2,\Q}+B_{2,\Q})C_{0,\K}\right]\nonumber\\
&\times\left(\frac{1}{\hbar\left(\omega-\omega^{(1)}_{2,\Q}-\omega^{(1)}_{0,\K}\right)+i\eta}-\frac{1}{\hbar\left(\omega+\omega^{(1)}_{2,\Q}+\omega^{(1)}_{0,\K}\right)-i\eta}\right)+\frac{4n_0c_2^2}{25}\integralQ C_{2,\Q}C_{0,\K}\nonumber\\
&\times\left(\frac{1}{\hbar\left(\omega-\omega^{(1)}_{2,\Q}-\omega^{(1)}_{0,\K}\right)+i\eta}-\frac{1}{\hbar\left(\omega+\omega^{(1)}_{2,\Q}+\omega^{(1)}_{0,\K}\right)-i\eta}\right)+c_0\integralQ \left(-C_{2,\Q}+\frac{c_2n_0}{10\eps{\Q}}\right)\nonumber\\
&+2c_1\integralQ \left[-C_{1,\Q}+\frac{(3c_1-c_2/5)n_0}{2\eps{\Q}}\right]-4c_1\integralQ \left(-C_{2,\Q}+\frac{c_2n_0}{10\eps{\Q}}\right)+\frac{c_2}{5}\integralQ \nonumber\\
&\times\left\{2\left(-C_{2,\Q}+\frac{c_2n_0}{10\eps{\Q}}\right)-2\left[-C_{1,\Q}+\frac{(3c_1-c_2/5)n_0}{2\eps{\Q}}\right]+\left[-C_{0,\Q}+\frac{(c_0+c_2/5)n_0}{2\eps{\Q}}\right]\right\},
\label{eq: summing Feynman diagrams for Sigma 12(2), 2,-2 for UN phase}
\end{align}

\begin{align}
\mu^{(2)}=&2c_0\integralQ\left(B_{2,\Q}+B_{1,\Q}+B_{0,\Q}\right)+6c_1\integralQ B_{1,\Q}+\frac{2c_2}{5}\integralQ B_{0,\Q}\nonumber\\
&+c_0\integralQ \left[-C_{0,\Q}+\frac{(c_0+c_2/5)n_0}{2\eps{\Q}}\right]+6c_1\integralQ \left[-C_{1,\Q}+\frac{(3c_1-c_2/5)n_0}{2\eps{\Q}}\right]\nonumber\\
&+\frac{c_2}{5}\integralQ\left\{2\left[-C_{2,\Q}+\frac{c_2n_0}{10\eps{\Q}}\right]-2\left[-C_{1,\Q}+\frac{(3c_1-c_2/5)n_0}{2\eps{\Q}}\right]+\left[-C_{0,\Q}+\frac{(c_0+c_2/5)n_0}{2\eps{\Q}}\right]\right\}.
\label{eq: summing Feynman diagrams for mu(2) for UN phase}
\end{align}
\end{widetext}
Since the emergent energy gap of the qNG modes $\hbar\omega_{\pm2,\p=\boldsymbol{0}}$ is smaller than $|c_1|n_0$ by a factor which is a function of the parameter $na^3\ll 1$, we can make Taylor series expansions of $\Sigma^{11(2)}_{22}$, $\Sigma^{22(2)}_{22}$, and $\Sigma^{12(2)}_{2,-2}$ in powers of $\hbar\omega_{\pm2,\p=\boldsymbol{0}}/(|c_1|n_0)$ and ignore the quadratic and higher-order terms. This is justified \textit{a posteriori} by the final result of the energy gap [Eq.~\eqref{eq: energy gap of quasi-NG modes}]. The second-order self-energies and chemical potential are then evaluated straightforwardly, and we obtain
\begin{widetext}
\begin{align}
\frac{\hbar^4\Sigma^{11(2)}_{22}(\omega_{\pm2, \p=\boldsymbol{0}},\p=\boldsymbol{0})}{M^{3/2}}=&\frac{n_0c_0^2}{\pi^2}\sqrt{n_0\tilde{c}_0}+\frac{12n_0c_1^2}{\pi^2}\sqrt{3n_0\tilde{c}_1}+\frac{n_0c_0}{\pi^2}\sqrt{n_0\tilde{c}_2^3}+\frac{2n_0c_0}{3\pi^2}\sqrt{n_0(3\tilde{c}_1)^3}+\frac{n_0c_0}{3\pi^2}\sqrt{n_0\tilde{c}_0^3}\nonumber\\
&+\frac{2n_0c_1}{3\pi^2}\sqrt{n_0(3\tilde{c}_1)^3}+\frac{4n_0c_1}{3\pi^2}\sqrt{n_0\tilde{c}_2^3}+\frac{2n_0c_2}{15\pi^2}\sqrt{n_0\tilde{c}_2^3}+\frac{3\sqrt{2}n_0c_1^2}{\pi^2}\Bigg[\sqrt{6n_0\tilde{c}_1}\nonumber\\
&-\frac{1}{\sqrt{6n_0\tilde{c}_1}}\hbar\omega_{\pm2, \p=\boldsymbol{0}}\Bigg]+\frac{n_0c_0^2}{\sqrt{2}\pi^2}\Bigg\{\frac{\sqrt{10}n_0^{1/2}\left[5c_0\sqrt{5\tilde{c}_0}+c_2\left(\sqrt{5\tilde{c}_0}+\sqrt{5\tilde{c}_2}\right)\right]}{75c_0+30c_2}\nonumber\\
&-\frac{\sqrt{10}\left[5c_0\sqrt{5\tilde{c}_0}+4c_2\sqrt{5\tilde{c}_0}+2(5\tilde{c}_2)^{3/2}\right]}{3(5c_0+2c_2)^2n_0^{1/2}}\hbar\omega_{\pm2, \p=\boldsymbol{0}}\Bigg\}\nonumber\\
&+\frac{2\sqrt{2}n_0c_0c_2}{5\pi^2}\Bigg\{\frac{c_2n_0^{1/2}}{\sqrt{10}(\sqrt{5\tilde{c}_2}+\sqrt{5\tilde{c}_0})}+\Bigg(\frac{\tilde{c}_2\left[(\sqrt{\tilde{c}_0}+\sqrt{\tilde{c}_2})^2\ln\left(\frac{\tilde{c}_0}{\tilde{c}_2}\right)-4(\tilde{c}_0-\tilde{c}_2)\right]}{4\sqrt{2}(\sqrt{\tilde{c}_0}+\sqrt{\tilde{c}_2})(\tilde{c}_0-\tilde{c}_2)^2n_0}\nonumber\\
&-\frac{n_0\tilde{c}_2\sqrt{\tilde{c}_0}}{\sqrt{\tilde{c}_0}+\sqrt{\tilde{c}_2}}\alpha\Bigg)\hbar\omega_{\pm2, \p=\boldsymbol{0}}\Bigg\}+\frac{2\sqrt{2}n_0c_2^2}{25\pi^2}\Bigg\{-\frac{\sqrt{n_0}(\sqrt{\tilde{c}_0}-\sqrt{\tilde{c}_2})^2}{3\sqrt{2}(\sqrt{\tilde{c}_0}+\sqrt{\tilde{c}_2})}-\tilde{c}_0\tilde{c}_2n_0^2\alpha\nonumber\\
&+\Bigg[\frac{3(\sqrt{\tilde{c}_0}+\sqrt{\tilde{c}_2})(\tilde{c}_0+\tilde{c}_2)\ln\left(\frac{\tilde{c}_0}{\tilde{c}_2}\right)-8\left(2\tilde{c}_0^{3/2}-3\tilde{c}_0\tilde{c}^{1/2}+3\tilde{c}_0^{1/2}\tilde{c}_2-2\tilde{c}_2^{3/2}\right)}{12\sqrt{2}(\tilde{c}_0-\tilde{c}_2)^2n_0^{1/2}}\nonumber\\
&+\frac{n_0\sqrt{\tilde{c}_0\tilde{c}_2}(\sqrt{\tilde{c}_0}-\sqrt{\tilde{c}_2})}{\sqrt{\tilde{c}_0}+\sqrt{\tilde{c}_2}}\alpha\Bigg]\hbar\omega_{\pm2, \p=\boldsymbol{0}}\Bigg\},
\label{eq: evaluate Sigma 11(2), 22 in the appendix}
\end{align}
where $\tilde{c}_0\equiv c_0+c_2/5, \tilde{c}_1\equiv c_1-c_2/15, \tilde{c}_2\equiv -c_2/5$, and 
\begin{align}
\alpha\equiv\frac{1}{n_0^{3/2}}\int_0^\infty\mathrm{d}x\frac{1}{2x\sqrt{(x+2\tilde{c}_0)(x+2\tilde{c}_2)}(\sqrt{x+2\tilde{c}_0}+\sqrt{x+2\tilde{c}_2})}. 
\end{align}
Despite $\alpha$ being infrared divergent, which is characteristic of the Bose-Einstein condensation, it does not affect the final result as shown below. Similarly, we have
\begin{align}
\frac{\hbar^4\Sigma^{12(2)}_{2,-2}(\omega_{\pm2, \p=\boldsymbol{0}},\p=\boldsymbol{0})}{M^{3/2}}=&\frac{3\sqrt{2}n_0c_1^2}{\pi^2}\sqrt{6\tilde{c}_1n_0}+\frac{n_0^{3/2}c_0^2c_2}{5\pi^2(\sqrt{\tilde{c}_2}+\sqrt{\tilde{c}_0})}-\frac{c_0(\tilde{c}_2n_0)^{3/2}}{\pi^2}+\frac{2c_1(3\tilde{c}_1n_0)^{3/2}}{\pi^2}+\frac{4c_1(\tilde{c}_2n_0)^{3/2}}{\pi^2}\nonumber\\
&+\frac{c_2}{5\pi^2}\left[-2(\tilde{c}_2n_0)^{3/2}-2(3\tilde{c}_1n_0)^{3/2}+(\tilde{c}_0n_0)^{3/2}\right]+\frac{2\sqrt{2}n_0^3c_2^2\tilde{c}_2\tilde{c}_0}{25\pi^2}\alpha\nonumber\\
&+\frac{2c_0c_2}{5\pi^2}\frac{\left[10c_0n_0\sqrt{\tilde{c}_0n_0}+5c_2n_0\sqrt{\tilde{c}_0n_0}+(5\tilde{c}_2n_0)^{3/2}\right]}{15c_0+6c_2},
\label{eq: evaluate Sigma 12(2), 22 in the appendix}
\end{align}
and
\begin{align}
\frac{\hbar^3\mu^{(2)}}{M^{3/2}}=&\frac{2c_0n_0}{3\pi^2}\left(\sqrt{n_0\tilde{c}_2^3}+\sqrt{n_0(3\tilde{c}_1)^3}+\sqrt{n_0\tilde{c}_0^3}\right)+\frac{2c_1n_0}{\pi^2}\sqrt{n_0(3\tilde{c}_1)^3}+\frac{2c_2n_0}{15\pi^2}\sqrt{n_0\tilde{c}_0^3}+\frac{c_0(\tilde{c}_0n_0)^{3/2}}{\pi^2}+\frac{6c_1(3\tilde{c}_1n_0)^{3/2}}{\pi^2}\nonumber\\
&+\frac{c_2}{5\pi^2}\left[-2(\tilde{c}_2n_0)^{3/2}-2(3\tilde{c}_1n_0)^{3/2}+(\tilde{c}_0n_0)^{3/2}\right].
\label{eq: evaluate mu(2) in the appendix}
\end{align}
\end{widetext}
The above expressions for the second-order self-energies [Eqs.~\eqref{eq: evaluate Sigma 11(2), 22 in the appendix} and \eqref{eq: evaluate Sigma 12(2), 22 in the appendix}] can be written in a concise form as
\begin{align}
\hbar\Sigma^{11(2)}_{2,2}=&A+B\hbar\omega_{\pm2,\p=\boldsymbol{0}}+\mathcal{O}\left[\left(\frac{\hbar\omega_{\pm2,\p=\boldsymbol{0}}}{|c_1|n_0}\right)^2\right], \label{eq: expansion of Sigma11,22}\\
\hbar\Sigma^{22(2)}_{2,2}=&A-B\hbar\omega_{\pm2,\p=\boldsymbol{0}}+\mathcal{O}\left[\left(\frac{\hbar\omega_{\pm2,\p=\boldsymbol{0}}}{|c_1|n_0}\right)^2\right],\\
\hbar\Sigma^{12(2)}_{2,-2}=&C+\mathcal{O}\left[\left(\frac{\hbar\omega_{\pm2,\p=\boldsymbol{0}}}{|c_1|n_0}\right)^2\right].
\label{eq: expansion of Sigma12,2,-2}
\end{align}

Substituting Eqs.~\eqref{eq: expansion of Sigma11,22}-\eqref{eq: expansion of Sigma12,2,-2} into Eq.~\eqref{eq: solution to the Dyson equation: omega(2,p=0) for UN phase}, we obtain
\begin{align}
\hbar\omega_{\pm2,\p=\boldsymbol{0}}=&\,\frac{\sqrt{\left[-\frac{c_2n_0}{5}+A-\mu^{(2)}\right]^2-\left[\frac{c_2n_0}{5}+C\right]^2}}{1-B}.
\label{eq: expression for the pole of the second-order Green's function G22}
\end{align}
From Eqs.~\eqref{eq: evaluate Sigma 11(2), 22 in the appendix}--\eqref{eq: evaluate mu(2) in the appendix}, the coefficients in the expansions [Eqs.~\eqref{eq: expansion of Sigma11,22}--\eqref{eq: expansion of Sigma12,2,-2}] are given by
\begin{align}
\frac{A-\mu^{(2)}+C}{(Mn_0)^{3/2}}=&\,\frac{1}{\pi^2\hbar^3}\Bigg(8\sqrt{3}\tilde{c}_1^{5/2}-\frac{32}{\sqrt{3}}\tilde{c}_1^{3/2}\tilde{c}_2+\frac{16}{3}\tilde{c}_1\tilde{c}_2^{3/2}\nonumber\\
&+\frac{8}{\sqrt{3}}\tilde{c}_1^{1/2}\tilde{c}_2^2-\frac{16}{9}\tilde{c}_2^{5/2}\Bigg),\label{eq: coefficient A-mu+C (2)}\\
\frac{A-\mu^{(2)}-C}{(Mn_0)^{3/2}}=&\, -\frac{16\sqrt{3}c_1^{5/2}}{\pi^2\hbar^3}+\mathcal{O}\left(c_2n_0\sqrt{na^3}\right), \label{eq: coefficient A-mu-C}\\
B=&\,0+\mathcal{O}\left(\sqrt{na^3}\right). \label{eq: coefficient B (2)}
\end{align}
Here we have ignored terms involving the factor of $\sqrt{na^3}\ll1$. Substituting Eqs.~\eqref{eq: coefficient A-mu+C (2)}--\eqref{eq: coefficient B (2)} into Eq.~\eqref{eq: expression for the pole of the second-order Green's function G22}, we find the emergent energy gap of the qNG modes to be
\begin{align}
\Delta=&\,c_1n\Bigg\{\Bigg[8\sqrt{3}\left(\frac{\tilde{c}_1}{c_1}\right)^{\frac{5}{2}}-\frac{32}{\sqrt{3}}\left(\frac{\tilde{c}_1}{c_1}\right)^{\frac{3}{2}}\frac{\tilde{c}_2}{c_1}+\frac{16}{3}\frac{\tilde{c}_1}{c_1}\left(\frac{\tilde{c}_2}{c_1}\right)^{\frac{3}{2}}\nonumber\\
&+\frac{8}{\sqrt{3}}\left(\frac{\tilde{c}_1}{c_1}\right)^{\frac{1}{2}}\left(\frac{\tilde{c}_2}{c_1}\right)^2-\frac{16}{9}\left(\frac{\tilde{c}_2}{c_1}\right)^{\frac{5}{2}}\Bigg]\Bigg[\frac{2|c_2|}{5c_1}\nonumber\\
&-\frac{128\sqrt{3}}{\sqrt{\pi}}\left(\frac{c_1}{c_0}\right)^{\frac{3}{2}}\sqrt{na^3}\Bigg]\frac{8}{\sqrt{\pi}}\left(\frac{c_1}{c_0}\right)^{\frac{3}{2}}\sqrt{na^3}\Bigg\}^{\frac{1}{2}}.
\label{eq: energy gap of quasi-NG modes}
\end{align}
By introducing the ratio $x\equiv-c_2/(15c_1)$, the right-hand side of Eq.~\eqref{eq: energy gap of quasi-NG modes} can be rewritten as
\begin{align}
\left(\frac{\Delta}{c_1n}\right)^2=f(x)\left[x-\frac{64}{\sqrt{3\pi}}\left(\frac{c_1}{c_0}\right)^\frac{3}{2}\sqrt{na^3}\right]\left(\frac{c_1}{c_0}\right)^\frac{3}{2}\sqrt{na^3},
\label{eq: energy gap of quasi-NG modes 2}
\end{align}
where 
\begin{align}
f(x)\equiv& 64\sqrt{\frac{108}{\pi}}\Big[(1+x)^\frac{5}{2}-4x(1+x)^\frac{3}{2}+2x^\frac{3}{2}(1+x)\nonumber\\
&+3x^2(1+x)^\frac{1}{2}-2x^\frac{5}{2}\Big].
\end{align}
Thus, Eq. (3) in the main text has been derived. The plot of $\Delta$ as a function of the ratio $-c_2/c_1$ is shown in Fig. 2 of the main text.

In the limit of $-c_2\sim c_1 (c_1/c_0)^{3/2}\sqrt{na^3}\ll c_1$, $f(x)\simeq 64\sqrt{108/\pi}$, and the energy gap reduces to
\begin{align}
\frac{\Delta}{c_1n}\simeq&8\left(\frac{108}{\pi}\right)^\frac{1}{4}\left(na^3\right)^\frac{1}{4}\left(\frac{c_1}{c_0}\right)^\frac{3}{4}\sqrt{x-\frac{64}{\sqrt{3\pi}}\left(\frac{c_1}{c_0}\right)^\frac{3}{2}\sqrt{na^3}}.
\label{eq: energy gap of quasi-NG modes for small c2/c1}
\end{align}
In particular, the energy gap at the UN-cyclic phase boundary $c_2^\mathrm{UN-CL}\simeq-342c_1(c_1/c_0)^{3/2}\sqrt{na^3}$~\cite{Phuc13b} gives the lower bound for $\Delta$:
\begin{align} 
\Delta_\mathrm{min}\simeq&\,\frac{64\sqrt{2}\sqrt[4]{3}}{\sqrt{\pi}}\left(\frac{7\sqrt{3}-8\sqrt{2}}{5}\right)^\frac{1}{2}c_1n\left(\frac{c_1}{c_0}\right)^\frac{3}{2}\sqrt{na^3}\nonumber\\
\simeq&\,27.06\,c_1n\left(\frac{c_1}{c_0}\right)^\frac{3}{2}\sqrt{na^3}.
\label{eq: minimum possible energy gap}
\end{align}

On the other hand, for $c_2\simeq -c_1$ the second term in square brackets in Eq.~\eqref{eq: energy gap of quasi-NG modes 2} is negligible compared with the first one, and the energy gap reduces to
\begin{align}
\Delta\simeq&c_1n \left(na^3\right)^\frac{1}{4}\left(\frac{c_1}{c_0}\right)^\frac{3}{4}\sqrt{xf(x)}.
\end{align}
At $c_2=-c_1$, i.e., $x=1/15$, the value of the energy gap is given by $\Delta\simeq 48/(5\sqrt[4]{5\pi})c_1n(c_1/c_0)^{3/4}\sqrt[4]{na^3}$.

\section{Propagation velocity of the qNG modes}
\label{sec: Propagation velocity of the qNG modes}
To calculate the propagation velocity of the qNG modes, we need to derive their dispersion relation, i.e., the excitation spectrum at nonzero momenta. In the low-momentum regime, the Bogoliubov dispersion relation of the qNG modes [Eq.~\eqref{eq: first-order omega +-2}] is linear: $\omega^{(1)}_{\pm2,\p}=v^{(1)}_\mathrm{q-NG}|\p|$ with the first-order propagation velocity given by 
\begin{align}
v^{(1)}_\mathrm{qNG}=\sqrt{\frac{|c_2|n}{5M}}.
\end{align}
For the spin-2 $\Rb$ BEC, since $c_0\gg c_1, |c_2|$, the spin-independent interaction predominates over the spin-dependent interactions. Therefore, concerning the modification of the propagation velocity, we can concentrate on $c_0$, which corresponds to the effect of fluctuations in the particle-number density, and ignore that of the spin-density fluctuations caused by $c_1$ and $c_2$. In the momentum regime satisfying $\Delta\ll \eps{\p}\ll |c_2|n$, where the modified dispersion relation is linear, it follows from the solution of the Dyson equation that the spectrum of the qNG modes can be expressed in terms of the second-order self-energy as 
\begin{align}
\omega^{(2)}_{\pm2,\p}\simeq\,\omega^{(1)}_{\pm2,\p}+\frac{\Sigma^{11(2)}_{22}(p)-\Sigma^{11(2)}_{22}(-p)}{2}.
\label{eq: omega +-2 for finite p}
\end{align}
Here we have ignored terms involving small factors of $\Delta/\eps{\p}$ and $\eps{\p}/|c_2|n$. If we neglect terms involving $c_1$ and $c_2$, Eq.~\eqref{eq: summing Feynman diagrams for Sigma 11(2), 22 for UN phase} for $\Sigma^{11(2)}_{22}(p)$ reduces to
\begin{widetext}
\begin{align}
\hbar\Sigma^{11(2)}_{22}(\omega,\p)=&n_0c_0^2\integralQ \left[(A_{0,\K}+B_{0,\K}-2C_{0,\K})\left(\frac{A_{2,\Q}}{\hbar(\omega-\omega^{(1)}_{2,\Q}-\omega^{(1)}_{0,\K})+i\eta}-\frac{B_{2,\Q}}{\hbar(\omega+\omega^{(1)}_{2,\Q}+\omega^{(1)}_{0,\K})-i\eta}\right)\right.\nonumber\\
&\left.-\mathcal{P}\frac{1}{\eps{\p}-\eps{\Q}-\eps{\K}}\right]+c_0\integralQ B_{0,\Q}.
\label{eq: reduced sigma 11,22}
\end{align}
\end{widetext}
Since $|\omega^{(2)}_{\pm2,\p}/\omega^{(1)}_{\pm2,\p}-1|\ll 1$, which can be verified \textit{a posteriori} from the final result [Eq.~\eqref{eq: dispersion relation for nematons (2)}], it is appropriate to replace $\omega$ in Eq.~\eqref{eq: reduced sigma 11,22} by $\omega^{(1)}_{\pm2,\p}$. The self-energy $\Sigma^{11(2)}_{22}(\pm p)$ for low momenta can then be calculated straightforwardly by using a Taylor expansion in powers of $|\p|$, yielding
\begin{align}
\Sigma^{11(2)}_{22}(\pm p)=&\frac{5M^{3/2}n_0^{3/2}c_0^{5/2}}{3\pi^2\hbar^3}\mp\frac{8\sqrt{2}|c_2|^{1/2}n_0^{1/2}\sqrt{\eps{\p}}}{3\sqrt{5\pi}}\nonumber\\
&+\mathcal{O}(|\p|^2).
\label{eq: sigma 11,22 (+-p)}
\end{align}
Substituting Eq.~\eqref{eq: sigma 11,22 (+-p)} into Eq.~\eqref{eq: omega +-2 for finite p}, we obtain the second-order dispersion relations of the qNG modes:
\begin{align}
\hbar\omega^{(2)}_{\pm2,\p}\simeq&\left(1-\frac{8}{3\sqrt{\pi}}\sqrt{n_0a^3}\right)\sqrt{\frac{2|c_2|n_0\eps{\p}}{5}}\nonumber\\
=&\left(1-\frac{4}{\sqrt{\pi}}\sqrt{na^3}\right)\sqrt{\frac{2|c_2|n\eps{\p}}{5}}
\label{eq: dispersion relation for nematons (2)}
\end{align}
with the modified propagation velocity 
\begin{align}
v^{(2)}_\mathrm{q-NG}=\left(1-\frac{4\sqrt{na^3}}{\sqrt{\pi}}\right)\sqrt{\frac{|c_2|n}{5M}}.
\end{align}
Thus, Eq. (4) in the main text has been derived.
Here, in the last equality in Eq.~\eqref{eq: dispersion relation for nematons (2)}, we have used the fact that the condensate fraction is given by~\cite{Pethick-book} 
\begin{align}
\frac{n_0}{n}=1-\frac{8\sqrt{na^3}}{3\sqrt{\pi}}. 
\end{align}
